\documentclass[aps,prd,10pt,twocolumn,superscriptaddress,longbibliography,floatfix,preprintnumbers]{revtex4-2}

\usepackage{amsmath,amssymb,amsfonts}
\usepackage{graphicx}
\usepackage{xcolor}
\usepackage{hyperref}
\hypersetup{colorlinks=true,linkcolor=blue,citecolor=blue,urlcolor=blue}
\usepackage{booktabs}
\usepackage{siunitx}
\usepackage{multirow}
\usepackage{dcolumn}
\usepackage{ulem}
\graphicspath{{figures/}}

\newcommand{\Alc}{A_l^{(c)}}
\newcommand{\Al}{A_l}
\newcommand{\Aq}{A_q}
\newcommand{\Gzero}{G_0}
\newcommand{\tc}{t_c}

\newcommand{\sigPhi}{\sigma_\Phi}
\newcommand{\Msun}{M_\odot}
\newcommand{\ms}{\,\mathrm{ms}}

\newcommand{\Rthresh}{R_\mathrm{thresh}}

\begin{document}

\preprint{LA-UR-26-23220}

\title{An Analytic Threshold for LESA-Driven Negative ELN Flux Directions
in Core-Collapse Supernovae: Derivation and Population Census}

\author{Nicol\'as Viaux}
\email{nicolas.viaux@usm.cl}
\affiliation{Departamento de F\'isica, Universidad T\'ecnica Federico
Santa Mar\'ia, Casilla 110-V, Valpara\'iso, Chile}

\author{Lucas Johns}
\affiliation{Theoretical Division, Los Alamos National Laboratory, Los Alamos, New Mexico 87545, USA}

\date{\today}

\begin{abstract}
In a core-collapse supernova (CCSN), an electron-rich progenitor star
deleptonizes through the preferential emission of $\nu_e$ over
$\bar{\nu}_e$. Nevertheless, along some viewing directions the
energy-integrated emitted lepton-number flux may become negative because
of lepton-number emission self-sustained asymmetry (LESA). We derive a
simple diagnostic connecting LESA to the presence of negative
lepton-number-flux directions.
We validate the criterion against 33 independent three-dimensional CCSN simulations: 25 models from the Princeton/Fornax ensemble
($8.1$--$100\,\Msun$) and 8 models from the Garching ensemble including a
non-rotating, slow-rotating, and fast-rotating $15\,\Msun$ sequence.
Of 23 non-black-hole-forming Princeton models, 22 ($96\%$) cross the
threshold with median onset $\tc = 225\ms$ (IQR 162--264\,ms) and
cross-model scatter $\mathrm{CV} = 18.6\%$.
Full-sky flux-sign searches confirm that the analytic threshold
specifically identifies the transition at the anti-LESA pole, demonstrating
that our criterion captures a distinct, globally-driven transition rather than
early localized turbulent crossings.
The fast-rotating $15\,\Msun$ Garching model, in which rapid rotation
suppresses the LESA dipole, is correctly identified as a non-crosser
without any rotation parameter in the formula.
Both black-hole-forming Princeton models cross the threshold $\sim\!250\ms$
post-bounce and remain above threshold for $1807$ and $2463\ms$ before
collapse.
Given the prevalence of negative lepton-number-flux directions, the
energy flux of $\bar{\nu}_e$ from the next nearby CCSN may well exceed
that of $\nu_e$ along some lines of sight. Such directions are also
plausibly correlated with sustained neutrino fast flavor instability,
although establishing that connection would require local neutrino
phase-space distributions or a dedicated linear stability analysis.
Throughout this paper the relevant quantity is the
energy-integrated emitted flux field, i.e., a luminosity difference per
steradian rather than a neutrino number flux.
\end{abstract}

\maketitle

\section{Introduction}
\label{sec:intro}

Three-dimensional simulations show pronounced angular structure in the emitted electron lepton number
(ELN) flux during the post-bounce phase~\cite{Nagakura2021,GlasPRD2020,Abbar2021}.
Its origin is not uniform. Early, localized negative-flux patches
appear near the proto-neutron star (PNS) surface within the first
$\sim 100$--$200\ms$ post-bounce, driven by PNS convective turbulence at
small angular scales ($\ell \geq 3$--$4$). These features cover only a small
fraction of the sky at onset and require high-angular-resolution transport to
resolve. Later, global negative-flux regions extend over the hemisphere
opposite the large-scale dipolar lepton-number-emission asymmetry later known
as lepton-number emission self-sustained asymmetry (LESA)~\cite{TamborraApJ2014}, cover
$\sim 10$--$20\%$ of the sky, and are directly tied to the growth of that
dipole. No analytic rule has previously distinguished these two regimes or
predicted the timing of the global transition.

LESA is a spontaneous, large-scale dipolar asymmetry
in which net lepton number---the difference between $\nu_e$ and $\bar\nu_e$
emission---is radiated preferentially from one hemisphere of the PNS.
LESA is observed across a wide range of progenitor masses in multiple
simulation codes~\cite{Hanke2013,TamborraPRL2013,TamborraPRD2014,TamborraApJ2014,Janka2016,Burrows2020,oconnor2018exploring,glas2019effects,vartanyan2019temporal}. The anti-LESA
hemisphere, where the emitted $\bar\nu_e$ flux is locally enhanced relative
to $\nu_e$, is a natural candidate for a hemisphere-scale negative-ELN-flux
region.

The preferential emission of $\bar{\nu}_e$ over $\nu_e$ along some lines of sight is an important feature from an observational standpoint. Neutrino fast flavor instability (FFI) provides an additional motivation for studying
the angular structure of the electron lepton-number field in core-collapse
supernovae~\cite{Sawyer2016, volpe2024neutrinos, johns2025neutrino, tamborra2025neutrinos}. Spatial correlations between the LESA anti-pole and regions favorable
for flavor conversion were reported in Refs.~\cite{GlasPRD2020,Abbar2021},
but without deriving a predictive threshold for the large-scale flux geometry.

Here we derive that threshold.
We show that the condition
\begin{equation}
  \Alc(t) = \Gzero(t) + \Aq(t)
  \label{eq:threshold_intro}
\end{equation}
follows directly from the spherical harmonic expansion of the ELN flux
evaluated at the anti-LESA pole, and that it correctly identifies the onset
of the global LESA-driven flux sign change across 33 independent
three-dimensional simulations from two ensembles. The threshold is expressed
in $\ell \leq 2$ multipole amplitudes, making it a compact diagnostic for the
large-scale emission geometry.

The scope of the claim is important. The present paper does not test for
standard ELN crossings in the FFI sense, because the available simulation
products do not provide local neutrino phase-space distributions
$f(E,\hat{v},x)$ or the ingredients for a linear stability analysis. Our
result instead concerns the angular structure of the energy-integrated emitted
flux field. In the FFI
literature, the relevant quantity is the local ELN angular
distribution in phase space, not the energy-integrated emitted flux field
analyzed in this work. Throughout this paper we therefore use
``negative ELN-flux direction'' to mean a direction on the emitted sky where
the energy-integrated $\bar\nu_e$ flux exceeds the $\nu_e$ flux, and we do
not identify this condition with a standard ELN crossing in the FFI
sense. More precisely, the field analyzed below is the difference of
energy-integrated emitted luminosities per steradian, so all references to
``flux'' in this paper should be read in that sense rather than as a
neutrino number flux.

Key results of this paper:
\begin{enumerate}
\item Derivation of $\Alc = \Gzero + \Aq$ from first principles (Sec.~\ref{sec:derivation}).
\item Validation that the threshold identifies the anti-LESA-pole flux reversal
  specifically, not the earlier turbulent crossings at $\langle\theta\rangle\approx 99^\circ$
  (Sec.~\ref{sec:comparison}).
\item Population census of threshold crossing across 25 Princeton/Fornax models
  (Sec.~\ref{sec:princeton}).
\item Code-independent cross-validation with 8 Garching models, including
  a non-crosser in a fast-rotating progenitor
  (Sec.~\ref{sec:garching}).
\item Black-hole-forming models remain above threshold for $\sim 2\,\mathrm{s}$
  before collapse (Sec.~\ref{sec:bh}).
\end{enumerate}

\section{Simulation data and multipole decomposition}
\label{sec:data}

\subsection{Princeton/Fornax dataset}
\label{sec:princeton_data}

The primary dataset consists of 25 three-dimensional CCSN simulations from
the publicly available Princeton/Fornax
ensemble~\cite{Burrows2020,Vartanyan2019}, spanning progenitor masses from
$8.1$ to $100\,\Msun$, including two independent $9\,\Msun$ realizations
(\texttt{9a} and \texttt{9b}).
Simulations run for $0.4$--$8.5\,\mathrm{s}$ post-bounce.
The data products used here are the energy-integrated spherical harmonic (SH)
coefficients $a_{lm}(t)$ of the $\nu_e$ and $\bar\nu_e$ luminosity fields,
summed over 12 logarithmically-spaced energy bins ($\approx 1$--$226\,\mathrm{MeV}$),
evaluated on a $128\times256$ ($\theta\times\phi$) angular grid using the
simulation solid-angle weights from the companion grid file.
Two models collapse to black holes during the simulations
($12.25$ and $14\,\Msun$) and are treated separately in
Sec.~\ref{sec:bh}.
The two $9\,\Msun$ models (\texttt{9a} and \texttt{9b}) are independent
stochastic realizations of the same progenitor, providing an internal
consistency check on model-to-model variability.

\subsection{Garching dataset}
\label{sec:garching_data}

The secondary dataset consists of 8 three-dimensional simulations from
the Garching/Prometheus-Vertex program, with data obtained from the
Garching Core-Collapse Supernova Data Archive~\cite{GarchingArchive}
(\url{https://wwwmpa.mpa-garching.mpg.de/ccsnarchive/}).
Models span masses $11.2$, $15$ (NR, SR, and FR variants), $20$, $27$,
$40$, and $75\,\Msun$.
The model provenance is as follows.
The $11.2$, $20$, and $27\,\Msun$ observer-projection data come from
the early three-dimensional PROMETHEUS-VERTEX models analyzed in
Refs.~\cite{TamborraPRL2013,TamborraPRD2014,TamborraApJ2014}, with the
$27\,\Msun$ case published in Ref.~\cite{Hanke2013} and the
$20\,\Msun$ case in Ref.~\cite{Melson2015}; this model set is also
reviewed in Ref.~\cite{Janka2016}.
The $15\,\Msun$ NR/SR/FR rotation sequence corresponds to the models of
Ref.~\cite{Summa2018}, and the $40$ and $75\,\Msun$ black-hole-forming
models correspond to Ref.~\cite{Walk2020}.
SH coefficients are provided on an $88\times176$ angular grid.
The $15\,\Msun$ FR model has a shellular rotation profile
($\Omega = 2\pi\times0.5\,\mathrm{rad\,s}^{-1}$ at the iron core) and is
the central test case for rotation effects (Sec.~\ref{sec:garching}).

\subsection{Multipole amplitude extraction}
\label{sec:multipoles}

For each timestep $t$ we form the ELN SH coefficients:
\begin{equation}
  a_{lm}^\mathrm{ELN}(t) = a_{lm}^{\nu_e}(t) - a_{lm}^{\bar\nu_e}(t),
  \label{eq:aELN}
\end{equation}
and define the multipole amplitudes used throughout this work:
\begin{align}
  \Gzero(t)  &\equiv a_{00}^\mathrm{ELN}(t)/\sqrt{4\pi},
  \label{eq:G0def} \\
  \Al(t)     &\equiv \Bigl[\textstyle\sum_{m=-1}^{1}
               (a_{1m}^\mathrm{ELN})^2\Bigr]^{1/2},
  \label{eq:Aldef} \\
  \Aq(t)     &\equiv \Bigl[\textstyle\sum_{m=-2}^{2}
               (a_{2m}^\mathrm{ELN})^2\Bigr]^{1/2},
  \label{eq:Aqdef} \\
  \sigPhi(t) &\equiv \bigl[\Al^2 + \Aq^2 + A_3^2 + A_4^2\bigr]^{1/2}.
  \label{eq:sigmadef}
\end{align}
Real-valued SH normalization with the Condon-Shortley phase convention is
used throughout.
$\Gzero$ is the angle-averaged ELN luminosity per steradian (positive when
$\nu_e$ globally exceeds $\bar\nu_e$); $\Al$ and $\Aq$ are the rms
amplitudes of the $\ell=1$ and $\ell=2$ modes.

The dimensionless LESA ratio $\varepsilon(t) \equiv \Al(t)/\Gzero(t)$
quantifies the strength of the dipolar asymmetry relative to the
isotropic background.
In all crossing models, $\varepsilon$ grows from $\sim10^{-5}$ at early
post-bounce times to order unity near $\tc$ and can exceed $10^3$ at
late times as $\Gzero$ decreases while $\Al$ remains large.

\section{Analytic threshold derivation}
\label{sec:derivation}

\subsection{ELN flux and LESA reference frame}

The ELN flux field is:
\begin{equation}
  \Phi(\hat{n},t) = L_{\nu_e}(\hat{n},t) - L_{\bar\nu_e}(\hat{n},t),
  \label{eq:Phi}
\end{equation}
where $L_\nu(\hat{n},t)$ is the energy-integrated luminosity per steradian
in direction $\hat{n}$.
We decompose $\Phi$ in real spherical harmonics:
\begin{equation}
  \Phi(\hat{n},t)
  = \sum_{l=0}^{\infty}\sum_{m=-l}^{l} a_{lm}(t)\,Y_{lm}(\hat{n}).
  \label{eq:SH}
\end{equation}
The LESA reference frame is defined by aligning the $z$-axis with the
instantaneous dipole vector:
\begin{equation}
  \hat{n}_\mathrm{LESA}(t)
  \equiv \vec{a}_{1m}(t) / |\vec{a}_{1m}(t)|,
  \label{eq:LESA_axis}
\end{equation}
where $\vec{a}_{1m}$ denotes the $\ell=1$ SH coefficient vector.
In this frame, by construction:
\begin{equation}
  a_{10} = \Al\sqrt{4\pi/3}, \qquad a_{1,\pm1} = 0.
  \label{eq:lesa_frame_l1}
\end{equation}
For the quadrupole, three-dimensional simulations show that the LESA dipole
and the dominant $\ell=2$ mode share an approximate common
symmetry axis, so the $m=0$ component carries most of the quadrupole power:
\begin{equation}
  a_{20} \approx \Aq\sqrt{4\pi/5}, \qquad
  |a_{2,m\neq0}| \ll |a_{20}|.
  \label{eq:lesa_frame_l2}
\end{equation}
The validity of Eq.~(\ref{eq:lesa_frame_l2}) is confirmed by the quadrupole
alignment analysis of Sec.~\ref{sec:comparison}.

\subsection{Pole expansion and threshold}

Evaluating Eq.~(\ref{eq:SH}) at the anti-LESA pole
$\hat{n} = -\hat{z}$ ($\theta = \pi$):
\begin{align}
  Y_{l0}(\pi,\phi) &= \sqrt{\tfrac{2l+1}{4\pi}}\,P_l(-1)
                    = \sqrt{\tfrac{2l+1}{4\pi}}\,(-1)^l,
  \label{eq:Yl0pole}
\end{align}
where $P_l(-1) = (-1)^l$.
Odd-$\ell$ modes contribute negatively and even-$\ell$ modes positively
at the anti-LESA pole.
Substituting the LESA-frame amplitudes:
\begin{align}
  \Phi(\pi,t) &= \Gzero
    - \underbrace{a_{10}\cdot Y_{10}(\pi)}_{=\,\Al}
    + \underbrace{a_{20}\cdot Y_{20}(\pi)}_{=\,\Aq}
    + \mathcal{O}(A_3,A_4,\ldots)
  \nonumber\\
  &= \Gzero - \Al + \Aq + \mathcal{O}(A_3,A_4,\ldots).
  \label{eq:pole}
\end{align}
The condition $\Phi(\pi,\tc) = 0$---the ELN flux first reaches zero at the
anti-LESA pole---gives the threshold:
\begin{equation}
  \boxed{\Alc(t) = \Gzero(t) + \Aq(t).}
  \label{eq:threshold}
\end{equation}
The excess above threshold,
\begin{equation}
  \delta(t) \equiv \Al(t) - \Alc(t),
  \label{eq:delta}
\end{equation}
is positive when the anti-LESA pole has negative ELN flux.

The physical content of Eq.~(\ref{eq:threshold}) is transparent.
The ELN monopole $\Gzero > 0$ represents the global excess of $\nu_e$
over $\bar\nu_e$, which the dipole $\Al$ must overcome to produce a sign
change.
The quadrupole $\Aq > 0$ adds to the anti-LESA pole flux via
$P_2(-1) = +1$, so it acts as an additional barrier: larger $\Al$ is
required to produce a crossing when quadrupole deformation is present.
Conversely, larger $\Aq$ delays the onset of LESA-driven negative ELN flux.
This has a direct consequence for rotating progenitors where rotation
suppresses $\Al$ while potentially leaving $\Aq/\Gzero$ unchanged
(Sec.~\ref{sec:garching}).

\subsection{ELN sky map: spatial confirmation of the pole approximation}

\begin{figure}[htbp]
  \centering
  \includegraphics[width=\columnwidth]{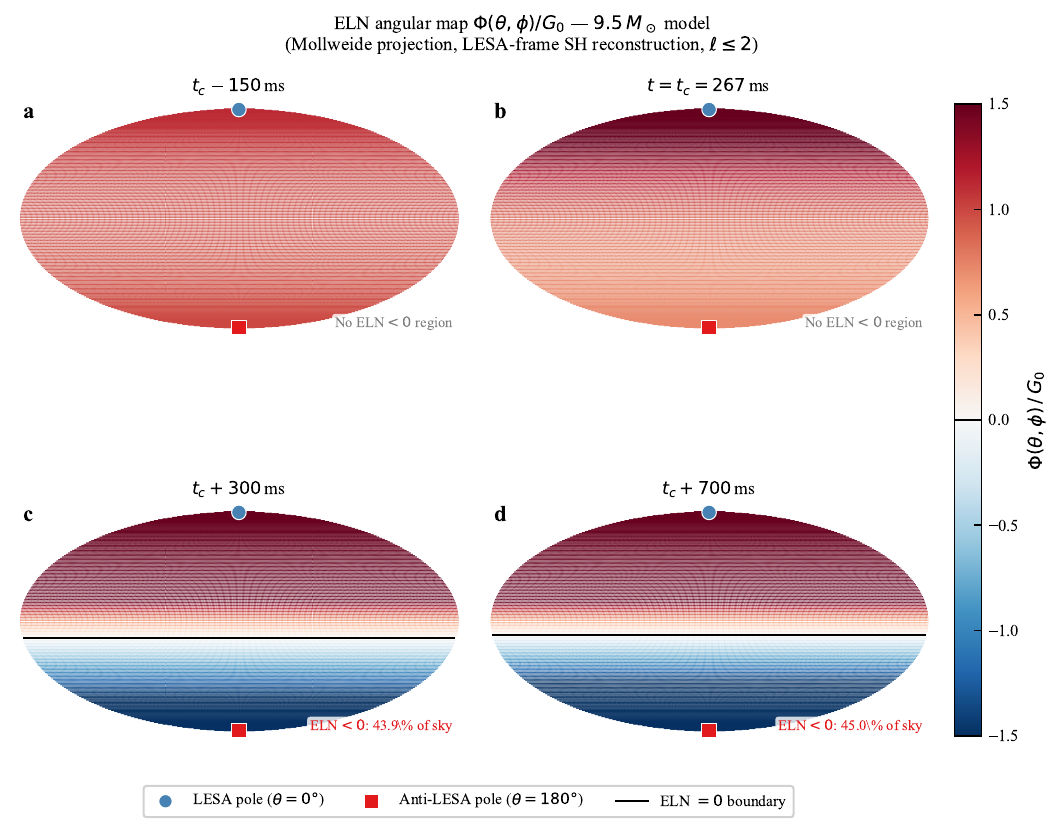}
  \caption{%
    \textbf{ELN sky map at threshold.}
    Mollweide projection of $\Phi(\theta,\phi)/\Gzero$ reconstructed from
    the $\ell\leq2$ SH expansion (Eq.~\ref{eq:map}) in the LESA frame
    for the $9.5\,\Msun$ fiducial model.
    Blue (red) shading: ELN negative (positive).
    Solid contour: ELN $= 0$ boundary.
    Blue circle: LESA pole ($\theta = 0^\circ$).
    Red square: anti-LESA pole ($\theta = 180^\circ$).
    \textbf{(a)} $t_c - 150\ms$: entire sky positive, no negative-flux region.
    \textbf{(b)} $t = \tc = 267\ms$: ELN first touches zero precisely at
    the anti-LESA pole, confirming the pole approximation.
    \textbf{(c)} $\tc + 300\ms$: ELN-negative region grows around the
    anti-LESA pole; LESA pole and equator remain positive.
    \textbf{(d)} $\tc + 700\ms$: negative region covers
    $\sim\!15$--$20\%$ of the sky.
    In the axisymmetric $\ell\leq2$ approximation, the negative-flux
    region originates at $\theta = 180^\circ$, consistent with
    Eq.~(\ref{eq:threshold}); nonzero $m\neq0$ components distort the
    contour away from perfect axisymmetry but do not change the fact that
    the global LESA-driven onset is centered on the anti-LESA direction.
  }
  \label{fig:skymap}
\end{figure}

Figure~\ref{fig:skymap} shows the full two-dimensional ELN map
$\Phi(\theta,\phi)/\Gzero$ reconstructed from the $\ell\leq2$ SH expansion
in the LESA frame for the fiducial $9.5\,\Msun$ model at four epochs.
The map is defined by:
\begin{align}
  \Phi(\theta,\phi)/\Gzero
  &= 1 + \varepsilon\cos\theta
  + (\Aq/\Gzero)\,P_2(\cos\theta) \nonumber\\
  &\quad + \mathcal{O}(A_3,A_4,\ldots),
  \label{eq:map}
\end{align}
where $\varepsilon = \Al/\Gzero$ and $P_2(\cos\theta) = (3\cos^2\theta-1)/2$.
Before $\tc$ (panel~a), the entire map is positive: no negative ELN-flux
direction exists.
At $\tc$ (panel~b), the map first touches zero at the anti-LESA pole
($\theta = 180^\circ$, red square), validating the pole approximation.
After $\tc$ (panels c, d), a growing ELN-negative region (blue) develops
exclusively around the anti-LESA pole; the LESA pole and equatorial regions
remain positive throughout.
The sky fraction with $\Phi < 0$ grows from $0\%$ at $\tc$ to
$\sim\!10$--$20\%$ at $\tc + 700\ms$.

\section{Onset time algorithm}
\label{sec:algo}

The onset time $\tc$ is defined as the earliest $t > 30\ms$ at which
$\delta(t) = \Al(t) - \Alc(t) > 0$ is sustained for at least 5 consecutive
output timesteps ($\lesssim 50\ms$ depending on model cadence), ensuring
that a brief fluctuation above threshold is not mistakenly identified as onset.

A \textbf{physical validity guard} rejects any epoch where
$\Gzero(t) \leq 0$.
A negative ELN monopole indicates that the angle-averaged $\bar\nu_e$
emission globally exceeds $\nu_e$, a regime in which the threshold formula
Eq.~(\ref{eq:threshold}) does not apply: the sign of $\Gzero$ sets which
pole (LESA or anti-LESA) would host a potential crossing, and when
$\Gzero < 0$, the LESA pole, not the anti-LESA pole, hosts the crossing.
This guard is essential for correctly treating the fast-rotating Garching
model (Sec.~\ref{sec:garching}), which shows a nominal $\delta > 0$ only
after $\Gzero$ has turned negative. Because the PNS deleptonizes from
initially electron-rich conditions, $\Gzero < 0$ is not expected to be a
generic outcome and appears here only in the unusual late-time evolution
of the fast-rotating Garching model.

The threshold ratio at crossing,
\begin{equation}
  \Rthresh \equiv \Alc/\Gzero\big|_{\tc}
  = 1 + \Aq/\Gzero\big|_{\tc},
  \label{eq:Rthresh}
\end{equation}
measures the fractional elevation of the threshold above the monopole
due to the quadrupole correction.
$\Rthresh = 1$ would correspond to $\Aq = 0$ (pure dipole threshold);
the observed values $\Rthresh = 1.42\pm0.36$ reflect the typical
$\sim\!40\%$ upward correction from the quadrupole at crossing
(Sec.~\ref{sec:princeton}). In short, $\Rthresh$ is a direct measure of
how important the quadrupole is in elevating the crossing threshold above
the pure-dipole value.

\section{Princeton/Fornax population census}
\label{sec:princeton}

\subsection{Overall crossing statistics}
\label{sec:census}

\begin{figure}[htbp]
  \centering
  \includegraphics[width=\columnwidth]{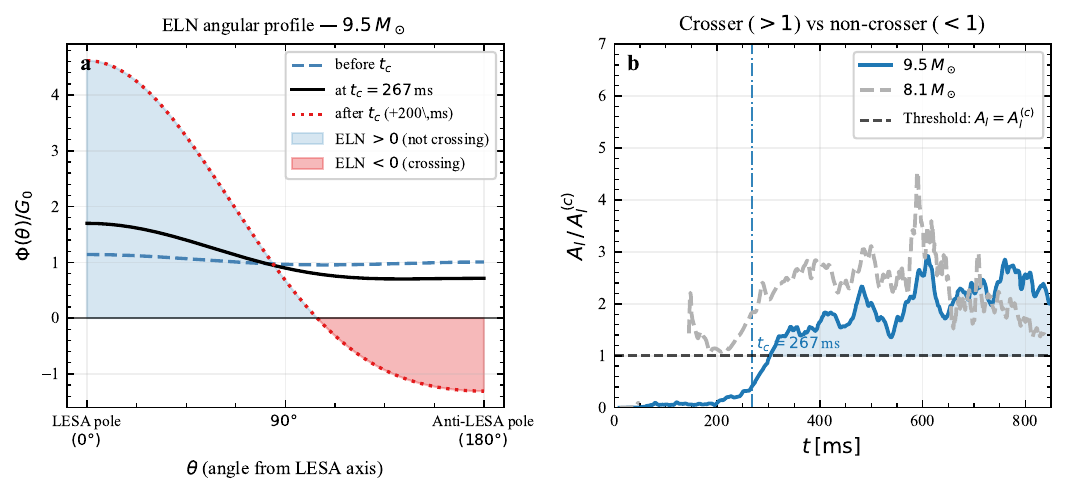}
  \caption{%
    \textbf{LESA geometry and analytic threshold for the
    $9.5\,\Msun$ fiducial model.}
    \textbf{(a)} Normalized ELN profile $\Phi(\theta)/\Gzero$ vs.\ angle
    from the LESA axis at three epochs:
    before $\tc$ (blue dashed), at $\tc = 267\ms$ (black solid), and
    $200\ms$ after $\tc$ (red dotted).
    The profile is axisymmetric in the LESA frame.
    The anti-LESA pole ($\theta = 180^\circ$) is the unique location
    where $\Phi$ first touches zero; the LESA pole ($\theta = 0^\circ$)
    remains positive throughout.
    Blue (red) shading marks ELN-positive (negative) regions.
    \textbf{(b)} Ratio $\Al(t)/\Alc(t)$ vs.\ time for the
    $9.5\,\Msun$ crossing model (blue) and the $8.1\,\Msun$ non-crossing
    model (grey dashed).
    The ratio crosses unity at $\tc = 267\ms$ for the $9.5\,\Msun$ model
    (vertical line); the blue-shaded area marks the interval with
    $\delta > 0$.
    The $8.1\,\Msun$ ratio remains below unity through the full
    $838\ms$ simulation, confirming that the threshold is not trivially
    satisfied.
  }
  \label{fig:fig1}
\end{figure}

Figure~\ref{fig:fig1} illustrates the threshold criterion for the
fiducial $9.5\,\Msun$ model.
Panel~(a) shows the ELN profile evolving from entirely positive (before
$\tc$) to developing a negative region at the anti-LESA pole after
$\tc = 267\ms$.
Panel~(b) shows the ratio $\Al/\Alc$ crossing unity at $\tc$ for the
$9.5\,\Msun$ model while remaining below unity throughout for the
$8.1\,\Msun$ model---the sole Princeton non-crosser.

\begin{figure*}[htbp]
  \centering
  \includegraphics[width=\textwidth]{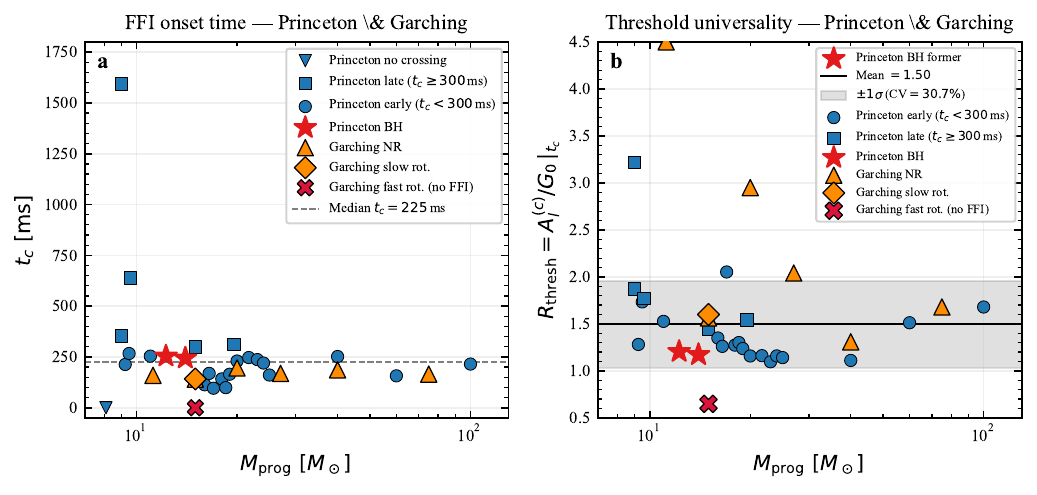}
  \caption{%
    \textbf{Population census of threshold crossing across Princeton and Garching
    ensembles.}
    \textbf{(a)} Threshold crossing time $\tc$ vs.\ progenitor mass for all 33
    simulations.
    Symbol scheme:
    blue circles = Princeton early crossers ($\tc < 300\ms$, $n=17$);
    blue squares = Princeton late crossers ($300\leq\tc < 600\ms$, $n=5$);
    grey downward triangle = Princeton non-crosser ($8.1\,\Msun$, plotted
    at $\tc = 0$);
    red stars = Princeton BH-forming models ($12.25$ and $14\,\Msun$);
    orange upward triangles = Garching non-rotating models ($n = 6$);
    orange diamond = Garching slow-rotating model (15SR);
    crimson cross = Garching fast-rotating non-crosser (15FR, plotted
    below axis).
    Horizontal dashed line: median $\tc = 225\ms$ for Princeton
    non-BH crossers.
    Both ensembles are described by the same threshold formula despite
    being from different simulation suites.
    \textbf{(b)} Threshold ratio $\Rthresh = \Alc/\Gzero|_{\tc}$
    vs.\ progenitor mass.
    Horizontal solid line: mean $\Rthresh = 1.42$;
    grey band: $\pm1\sigma = \pm0.26$.
    Princeton and Garching models scatter around the same mean,
    confirming code-independent universality.
    Two physical outliers lie above the band: $9\,\Msun$ model
    \texttt{9b} ($\Rthresh = 3.22$, stochastic LESA variability) and
    $17\,\Msun$ ($\Rthresh = 2.05$, early crossing during PNS settling).
    The crimson cross (15FR) is plotted below the panel range at $y=0.65$,
    indicating no crossing.
    The Spearman anti-correlation between $\Aq/\Gzero$ and mass
    ($r = -0.58$, $p = 0.004$) is consistent with a weak tendency toward
    earlier onset at higher progenitor mass, albeit with substantial
    scatter.
  }
  \label{fig:census}
\end{figure*}

Table~\ref{tab:census} and Figure~\ref{fig:census} summarize the full
census.
Of the 23 non-BH Princeton progenitors, \textbf{22 cross the threshold}
($96\%$).
The sole non-crosser is the $8.1\,\Msun$ model (\texttt{u8.1}), in which
the LESA dipole $\Al$ grows too slowly relative to $\Gzero + \Aq$ before
the simulation ends at $838\ms$.
We note that the $8.1\,\Msun$ non-crossing verdict is
consistent with its classification in the literature as a very weakly
exploding model with subdued PNS convection.

Onset times span $\tc \in [96, 1594]\ms$:
\begin{itemize}
\item \textbf{Early crossers} ($\tc < 300\ms$): 17 models,
  median $\tc = 220\ms$.
\item \textbf{Late crossers} ($300 \leq \tc < 600\ms$): 5 models,
  median $\tc = 353\ms$.
\item \textbf{Non-crosser}: 1 model ($8.1\,\Msun$).
\item \textbf{BH formers}: 2 models, discussed in Sec.~\ref{sec:bh}.
\end{itemize}
Overall median $\tc = 225\ms$ with IQR $162$--$264\ms$ (combining all
non-BH crossers).

\begin{table*}[htbp]
\caption{Complete Princeton/Fornax threshold-crossing census.
$\tc$: onset time (ms). $\Rthresh$: threshold ratio at crossing.
$\varepsilon_\mathrm{peak}$: peak LESA ratio $\Al/\Gzero$.
$f_\mathrm{SASI}$: dominant oscillation frequency in $\Al(t)$ after
LESA saturation (Hz).
BH: black-hole-forming model.
NX: no crossing detected.}
\label{tab:census}
\begin{ruledtabular}
\begin{tabular}{lcccccl}
Label & $M$ ($\Msun$) & $\tc$ (ms) & $\Rthresh$ &
$\varepsilon_\mathrm{peak}$ & $f_\mathrm{SASI}$ (Hz) & Category \\
\hline
u8.1   &  8.1 &  --- & 1.61 &    137 & 20.2 & NX \\
9a     &  9.0 & 1594 & 1.87 &   5443 & 11.4 & late \\
9b     &  9.0 &  353 & 3.22 &    792 & 19.6 & late \\
9.25   &  9.25&  213 & 1.59 &  20617 & 20.7 & early \\
9.5    &  9.5 &  267 & 1.64 &    771 & 10.0 & early \\
z9.6   &  9.6 &  639 & 1.77 &     54 & 10.4 & late \\
11     & 11.0 &  254 & 1.22 &    130 & 16.6 & early \\
12.25  & 12.25&  254 & 1.06 &      4 & 14.3 & BH \\
14     & 14.0 &  244 & 1.29 &      7 & 10.0 & BH \\
15.01  & 15.0 &  300 & 1.25 &    431 & 17.4 & late \\
16     & 16.0 &  113 & 1.15 &   6018 & 16.8 & early \\
16.5   & 16.5 &  169 & 1.76 &    289 & 12.4 & early \\
17     & 17.0 &   96 & 1.21 &   2532 & 12.4 & early \\
18     & 18.0 &  142 & 1.12 &   1108 & 15.1 & early \\
18.5   & 18.5 &   99 & 1.17 &    996 & 11.7 & early \\
19     & 19.0 &  165 & 1.44 &   3052 & 10.2 & early \\
19.56  & 19.56&  311 & 1.60 &   1549 & 17.9 & late \\
20     & 20.0 &  230 & 1.16 &    554 & 13.3 & early \\
21.68  & 21.68&  247 & 1.46 &     20 & 14.3 & early \\
23     & 23.0 &  237 & 1.18 &  10705 & 12.9 & early \\
24     & 24.0 &  220 & 1.74 &   1536 & 10.9 & early \\
25     & 25.0 &  161 & 1.22 &    581 & 19.4 & early \\
40     & 40.0 &  252 & 1.37 &     27 & 10.2 & early \\
60     & 60.0 &  157 & 1.12 &   4376 & 11.5 & early \\
100    &100.0 &  215 & 1.30 &     79 &106.5 & early \\
\end{tabular}
\end{ruledtabular}
\end{table*}

\subsection{Threshold ratio universality}

The threshold ratio $\Rthresh$ (Fig.~\ref{fig:census}b) is:
\begin{equation}
  \Rthresh = 1.417 \pm 0.263 \quad (\mathrm{CV} = 18.6\%)
  \label{eq:Rthresh_census}
\end{equation}
for the 22 non-BH crossing models.
The minimum and maximum values are $\Rthresh = 1.119$ ($60\,\Msun$) and
$\Rthresh = 3.221$ ($9\,\Msun$~\texttt{9b}).
The two outliers both have clear physical explanations:
\begin{itemize}
\item \textbf{\texttt{9b}, $\Rthresh = 3.22$}:
  Stochastic variability between two independent realizations of the same
  $9\,\Msun$ progenitor produces a transient, anomalously large
  $\Aq/\Gzero$ episode.
  The companion model \texttt{9a} gives $\Rthresh = 1.87$, confirming
  that stochastic scatter between realizations of the same mass can span
  a factor of $\sim\!2$ in $\Rthresh$.

\item \textbf{$17\,\Msun$, $\Rthresh = 2.05$}:
  The earliest onset in the ensemble ($\tc = 96\ms$) occurs during the
  initial PNS settling phase, when both $\Gzero$ and $\Aq$ are rapidly
  evolving.
  A transient large-$\Aq/\Gzero$ ratio at this early epoch elevates
  $\Rthresh$ above the typical late-time value.
\end{itemize}
Excluding these two outliers, the remaining 20 models satisfy
$\Rthresh = 1.315\pm0.153$ ($\mathrm{CV} = 12\%$).

An anti-correlation between $\Aq/\Gzero$ and progenitor mass is detected
(Spearman $r = -0.58$, $p = 0.004$): more massive progenitors have
smaller relative quadrupole deformation at the time of crossing, lowering
$\Rthresh$ and contributing to a weak tendency toward earlier $\tc$ at higher
masses.

\subsection{LESA saturation times and SASI oscillations}

LESA saturation occurs when $\Al/\Gzero$ first exceeds $0.1$ and remains
elevated.
Saturation times range from $\sim 36\ms$ (Garching $15\,\Msun$ NR) to
$\sim 812\ms$ (Princeton \texttt{9a}).
In all models with $\tc$ detected, $\tc$ occurs well after LESA saturation,
confirming that onset requires LESA to be fully established, not merely
beginning. For models with $M \gtrsim 18\,\Msun$, $\Al(t)$ exhibits a quasi-periodic oscillation after saturation with frequencies $f_\mathrm{SASI} \approx
10$--$20\,\mathrm{Hz}$ (Table~\ref{tab:census}), consistent with SASI
modulation of the PNS accretion~\cite{TamborraPRL2013,TamborraPRD2014}.
The $100\,\Msun$ model shows an anomalously high frequency
($f = 106\,\mathrm{Hz}$) due to its rapid accretion timescale.

\section{Code-independent validation and rotation effects}
\label{sec:garching}

\begin{center}
  \refstepcounter{figure}
  \includegraphics[width=\columnwidth]{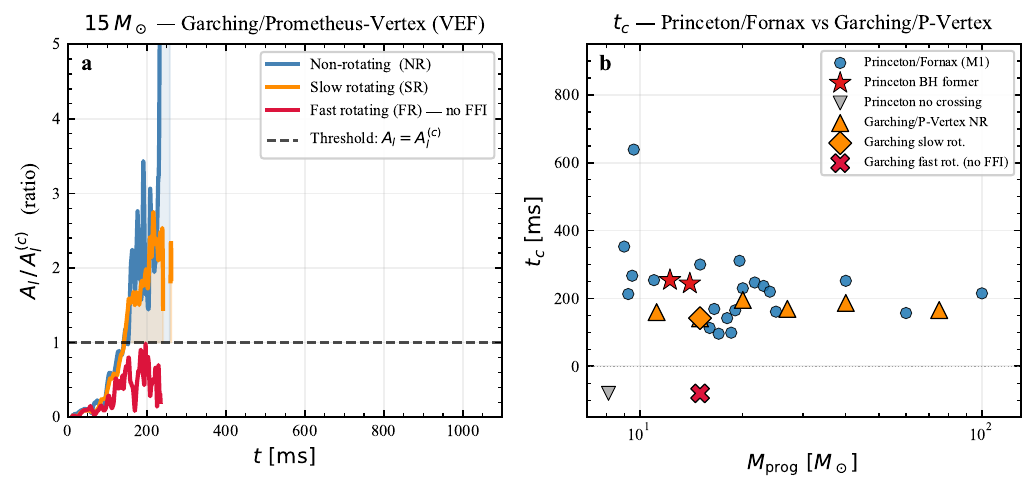}

  \small
  \textbf{Figure \thefigure. Garching validation and rotation sequence.}
  \textbf{(a)} Ratio $\Al(t)/\Alc(t)$ for the $15\,\Msun$ Garching
  rotation sequence: non-rotating NR (steelblue solid), slow-rotating SR
  (orange solid), fast-rotating FR (crimson solid).
  Shaded regions: epochs with $\delta > 0$ (ratio $> 1$).
  NR and SR cross the threshold at $\tc \approx 141$ and $142\ms$
  respectively; their onset times are nearly identical despite SR
  having $\sim50\%$ lower peak LESA amplitude, because $\Gzero + \Aq$
  is also suppressed proportionally.
  FR never exceeds unity while $\Gzero > 0$: rotation suppresses
  $\Al/\Gzero|_\mathrm{peak}$ from $\sim\!1000$ (NR) to $12.9$ (FR),
  preventing the threshold from being reached.
  The $\Gzero > 0$ validity region is shaded.
  \textbf{(b)} Cross-code comparison of $\tc$ vs.\ progenitor mass for
  all Princeton crossers (blue) and all Garching crossers (orange).
  Red stars: Princeton BH-forming models.
  Crimson cross: 15FR non-crosser, plotted below the axis.
  Grey triangle: Princeton non-crosser ($8.1\,\Msun$).
  Both simulation suites are described by the same threshold formula,
  and Garching $\tc$ values (141--196\,ms) overlap with Princeton onset
  times at comparable masses.
  \label{fig:garching}
\end{center}

\subsection{Non-rotating and slow-rotating Garching models}

Seven of the eight Garching models cross the threshold while $\Gzero > 0$,
with $\tc$ in the range $142$--$196\ms$ (Table~\ref{tab:garching}).
Figure~\ref{fig:garching}a shows the $15\,\Msun$ rotation sequence.
The NR and SR models cross at $\tc \approx 141$ and $142\ms$
respectively---a difference of only $1\ms$ despite SR having a
$\sim\!35\%$ lower peak LESA amplitude.
This near-degeneracy occurs because slow rotation suppresses both $\Al$
and $\Gzero + \Aq$ in similar proportion, leaving the ratio
$\Al/(\Gzero+\Aq)$ largely unchanged.
Figure~\ref{fig:garching}b and Fig.~\ref{fig:census}b confirm that
Garching $\Rthresh$ values scatter around the same mean ($1.42$) as
Princeton models.

\begin{center}
\refstepcounter{table}
\textbf{Table \thetable. Garching/Prometheus-Vertex threshold census.}
$t_\mathrm{sat}$: LESA saturation time (ms).
$\tc$: analytic threshold onset (ms).
$\varepsilon_\mathrm{peak} = \Al/\Gzero|_\mathrm{peak}$.
\label{tab:garching}

\begin{tabular}{lccccl}
\toprule
Label & $M$ ($\Msun$) & $t_\mathrm{sat}$ & $\tc$ &
$\varepsilon_\mathrm{peak}$ & Verdict \\
\midrule
11.2  & 11.2 &  56 & 160 &  2023 & crosses \\
15NR  & 15.0 &  37 & 141 &   995 & crosses \\
15SR  & 15.0 &  72 & 142 &   699 & crosses \\
\textbf{15FR}  & \textbf{15.0} & \textbf{46} & \textbf{---} &
  \textbf{12.9} & \textbf{non-crosser} \\
20    & 20.0 &  66 & 196 & 10056 & crosses \\
27    & 27.0 &  72 & 169 & 20710 & crosses \\
40    & 40.0 &  83 & 187 &   612 & crosses \\
75    & 75.0 & 143 & 166 &    85 & crosses \\
\bottomrule
\end{tabular}
\end{center}

\subsection{The fast-rotating non-crosser}
\label{sec:FR}

The $15\,\Msun$ FR model is the central validation test.
In this model, rapid shellular rotation ($\Omega = 2\pi\times0.5\,\mathrm{rad\,s}^{-1}$)
suppresses the LESA dipole to
$\Al/\Gzero|_\mathrm{peak} = 12.9$---a factor of $\sim\!80$ below the
NR value---while leaving the quadrupole correction $\Aq/\Gzero$ at
order unity.
As a result, $\Al$ never reaches $\Gzero + \Aq$ before $\Gzero$ itself
turns negative (Fig.~\ref{fig:garching}a, crimson curve), and the
physical validity guard (Sec.~\ref{sec:algo}) correctly classifies this
model as a non-crosser.

This is a useful non-crosser: a three-dimensional simulation that is
correctly rejected by the threshold formula, demonstrating
discriminatory power.
No rotation parameter appears in Eq.~(\ref{eq:threshold}): the effect of
rotation enters entirely through its observed modification of $\Al$ and
$\Aq/\Gzero$ in the data, making the criterion self-calibrating.

\section{Black-hole-forming models}
\label{sec:bh}

The two Princeton BH-forming models provide the most striking results
in the census.
Both cross the threshold well before collapse, with
$\tc = 254\ms$ ($12.25\,\Msun$) and $\tc = 244\ms$ ($14\,\Msun$),
consistent with non-BH-forming models of similar mass.
The low threshold ratios ($\Rthresh = 1.064$ and $1.295$) indicate minimal
quadrupole correction at crossing, consistent with the general trend of
lower $\Rthresh$ at higher mass.

What distinguishes the BH models is not the onset but the duration.
The intervals between threshold crossing and collapse are
$\Delta t_\mathrm{cross} = 1807\ms$ for the $12.25\,\Msun$ model and
$2463\ms$ for the $14\,\Msun$ model.
During these windows, $\delta(t) > 0$ continuously, so once the anti-LESA
pole flux turns negative it remains negative until collapse. The BH models
therefore stand out not because they cross earlier than the rest of the
ensemble, but because they sustain the LESA-driven large-scale negative-flux
configuration for $\sim 2\,\mathrm{s}$ before the neutrino emission cuts off.
\section{Comparison with full-sky negative-flux directions and higher-multipole
criteria}
\label{sec:comparison}

\begin{center}
  \refstepcounter{figure}
  \includegraphics[width=\columnwidth]{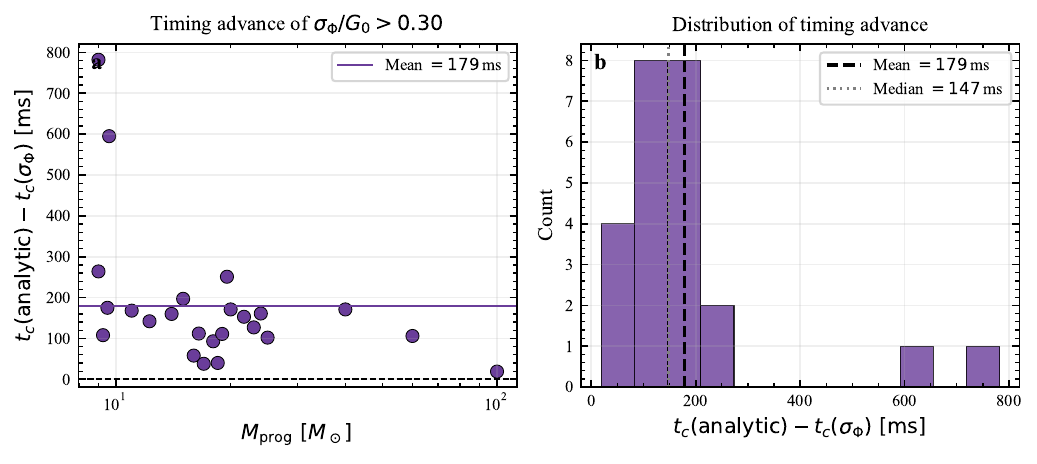}

  \small
  \textbf{Figure \thefigure. Comparison of threshold criteria.}
  \textbf{(a)} Timing of the $\sigPhi/\Gzero > 0.30$ empirical criterion
  (Eq.~\ref{eq:sigma}) vs.\ analytic threshold $\tc$ for all 25
  Princeton models.
  The empirical criterion fires $\sim\!419\ms$ earlier on average
  (points above the diagonal).
  Error bars reflect timestep cadence uncertainty.
  \textbf{(b)} Histogram of timing advance
  $\Delta t = \tc(\text{analytic}) - \tc(\sigPhi)$.
  Mean advance $419\ms$, caused by localized PNS convective negative-flux patches
  at $\ell \geq 3$ that the empirical criterion captures but the
  $\ell\leq2$ analytic threshold cannot.
  The analytic threshold predicts specifically the global, anti-LESA-pole
  flux reversal rather than these earlier turbulent events.
  \label{fig:sigma}
\end{center}

\subsection{Full-sky negative-flux analysis}

We performed a direct full-sky search for negative ELN-flux directions for all
25 Princeton models by computing the sign of $\Phi(\theta,\phi,t)$ on the
$128\times256$ angular grid at each timestep and identifying the first
epoch when any sky pixel sustains $\Phi < 0$ for multiple consecutive
timesteps.

Of 24 models with detectable negative-flux directions (all except
\texttt{u8.1}), the first negative-flux patch on the sky occurs
before the analytic threshold
in all cases, with mean advance
$\langle t_\mathrm{neg,first} - \tc\rangle = -210\ms$
(median $-152\ms$, standard deviation $305\ms$).
Crucially, the \textbf{mean angular position of the first negative-flux patch
is $\langle\theta_\mathrm{neg}\rangle = 99^\circ$}---near the equator,
not at the anti-LESA pole.
At first appearance, the sky fraction with $\Phi < 0$ is on average
only $12\%$---a localized, turbulence-driven feature at random sky
positions.

This result sharply distinguishes two types of negative-flux behavior:
\begin{enumerate}
\item \textbf{Early, localized negative-flux patches}
  ($\langle\theta\rangle = 99^\circ$, $\langle f_\mathrm{sky}\rangle = 12\%$,
  mean $210\ms$ before analytic $\tc$):
  driven by PNS convective turbulence at $\ell \geq 3$--$4$ scales,
  these features are stochastic, small-scale, and cannot be predicted
  from the $\ell\leq2$ emission geometry.

\item \textbf{Global, LESA-driven flux reversal}
  (at $\theta = 180^\circ$ by construction, onset predicted by
  $\Alc = \Gzero + \Aq$):
  driven by the large-scale LESA dipole, this transition grows from zero at
  the anti-LESA pole and expands to cover $\sim\!15$--$20\%$ of the sky
  within $\sim\!700\ms$ of $\tc$.
\end{enumerate}
Our analytic threshold predicts specifically the second type.
The 210\,ms temporal gap between the two types of behavior is a
quantitative measure of how much earlier small-scale turbulence
establishes localized negative-flux patches before the global LESA-driven transition
occurs.

\subsection{Empirical higher-multipole criterion}

The empirical criterion
\begin{equation}
  \sigPhi/\Gzero > 0.30,
  \quad \sigPhi = \bigl[\Al^2 + \Aq^2 + A_3^2 + A_4^2\bigr]^{1/2},
  \label{eq:sigma}
\end{equation}
achieves timing mean absolute error $\mathrm{MAE} = 10\ms$ relative to
the full-grid flux minimum (Fig.~\ref{fig:sigma}).
It fires $\sim\!419\ms$ before the analytic threshold on average, capturing
the turbulent $\ell\leq4$ negative-flux patches through the higher-multipole power
$A_3^2 + A_4^2$.

The two criteria are complementary:
\begin{itemize}
\item $\Alc = \Gzero + \Aq$ ($\ell\leq2$): predicts the global, anti-LESA-pole
  flux reversal from hemisphere-integrated observables alone.
  Provides the first onset computable from coarse angular resolution.
\item $\sigPhi/\Gzero > 0.30$ ($\ell\leq4$): achieves near-full-grid timing
  precision by incorporating higher-multipole turbulent structure.
  Requires $\ell=3,4$ transport output and does not identify the transition
  location.
\end{itemize}

\section{Discussion}
\label{sec:discussion}

\subsection{Relation to prior work}

Spatial correlations between the LESA anti-pole and ELN-structured emission regions
were established by Refs.~\cite{GlasPRD2020,Abbar2021}.
The present work provides the quantitative threshold explaining this
correlation: the global ELN sign-change at the anti-LESA pole first
occurs when $\Al = \Gzero + \Aq$.
The full-sky negative-flux analysis (Sec.~\ref{sec:comparison}) adds a key
clarification not present in those studies: earlier negative-flux patches at
$\langle\theta\rangle \approx 99^\circ$ are turbulent events unrelated to
the LESA dipole geometry.
Our threshold predicts the onset of the geometrically distinct, LESA-driven
flux reversal at $\theta = 180^\circ$.

Systematic surveys~\cite{Nagakura2021,Cornelius2025}
catalogued ELN angular-structure statistics across multiple simulations.
In particular, Ref.~\cite{Nagakura2021} distinguished type-I and
type-II crossings and discussed how turbulent crossings fit into that
taxonomy. The present work adds a complementary analytic threshold for the
later, large-scale LESA-driven transition and ties that transition
directly to the low-order multipole geometry.

Reference~\cite{Wang2025} demonstrated self-consistent FFI in
three-dimensional simulations.
The analytic threshold is complementary to such work: it provides a
transport-free, zero-cost diagnostic for screening simulation archives and
predicting the onset time of the large-scale anti-LESA flux reversal
directly from $\ell\leq2$ SH outputs.

\subsection{Validity of the \texorpdfstring{$\ell=2$}{l=2} pole approximation}

The approximation $a_{20} \approx \Aq\sqrt{4\pi/5}$ (dominant $m=0$
quadrupole in the LESA frame) was verified by computing the quadrupole
alignment angle $\psi$---the angle between the dipole axis
$\hat{n}_\mathrm{LESA}$ and the principal axis of the $\ell=2$ moment
tensor.
Across the Princeton ensemble, $\langle\psi\rangle \approx 15^\circ$
during the LESA saturation phase, confirming near-axisymmetric quadrupole
orientation and validating Eq.~(\ref{eq:lesa_frame_l2}).

Higher-order corrections ($A_3$, $A_4$) enter Eq.~(\ref{eq:pole}) with
alternating signs $(-1)^l$ and amplitudes $\sim A_l/\Gzero \ll 1$
during the LESA growth phase at $t \approx \tc$.
Non-axisymmetric $m\neq0$ terms can therefore shift the exact
minimum-flux direction slightly away from $\theta = 180^\circ$ on the full
sky, but for the globally driven transition they act as perturbations on
top of a dipole-plus-quadrupole geometry centered on the anti-LESA pole.
At late times ($t \gg \tc$), when $\varepsilon \gg 1$, the $\ell\leq2$
approximation overestimates $\Phi(\pi)$; however, the threshold crossing
itself occurs at early times and is not affected.

\subsection{18\% scatter: physics or numerics?}

The $\mathrm{CV} = 18.6\%$ scatter in $\Rthresh$ (Eq.~\ref{eq:Rthresh_census})
is driven by model-to-model variations in $\Aq/\Gzero$ at the moment
of crossing.
That both Princeton and Garching models scatter around the same mean
value ($1.42$) confirms that this scatter is a physical property of
CCSN convection variability, not an artifact of any particular simulation.
The two physical outliers (\texttt{9b} and $17\,\Msun$) are the primary
drivers of the elevated CV; excluding them gives $\mathrm{CV} = 12\%$.

\subsection{Implications for simulation archives}

The threshold formula and the $\sigPhi$ criterion (Eq.~\ref{eq:sigma})
provide complementary screening tools for negative-flux behavior.
The analytic threshold can be computed from any SH decomposition output
in seconds per model, without requiring full-grid angular resolution.
Applied to a simulation archive, it identifies models where global
LESA-driven negative flux appears and provides the onset time $\tc$ for each.
The $\sigPhi$ criterion (requiring $\ell\leq4$ output) provides a
more precise timing estimate for models where the higher-multipole
content is available.
Together they bracket the range from globally-predictable to
precisely-timed negative-flux onset.

\section{Conclusions}
\label{sec:conclusions}

We have derived and validated the first analytic threshold connecting
the LESA large-scale emission asymmetry to the onset of negative ELN flux
in core-collapse supernovae:
\begin{equation}
  \Alc(t) = \Gzero(t) + \Aq(t).
  \label{eq:threshold_conc}
\end{equation}
This criterion follows directly from the spherical harmonic expansion of
the ELN flux evaluated at the anti-LESA pole and identifies specifically
the onset of the global, LESA-driven flux reversal---distinct from
earlier, localized, turbulence-driven negative-flux patches at equatorial
sky positions.

Validation against 33 independent three-dimensional CCSN simulations
from two simulation ensembles establishes:
\begin{enumerate}
\item $96\%$ crossing rate (22/23 non-BH Princeton models),
  median $\tc = 225\ms$, cross-model scatter $\mathrm{CV} = 18.6\%$.
\item Full-sky flux-sign analysis confirms the predicted onset location
  ($\theta = 180^\circ$) vs.\ $\langle\theta\rangle = 99^\circ$ for
  earlier turbulent negative-flux patches.
\item Code-independent universality: Princeton ($\Rthresh = 1.42\pm0.26$)
  and Garching models scatter around the same mean.
\item Fast-rotating $15\,\Msun$ Garching model remains a non-crosser and is
  correctly rejected without a rotation parameter.
\item BH-forming models sustain the large-scale negative-flux state for
  $\sim\!2\,\mathrm{s}$ before abrupt signal cutoff.
\item The distinction between early turbulent patches and the later LESA-driven
  global transition can be captured with low-order multipole data alone.
\end{enumerate}

The criterion $\Alc = \Gzero + \Aq$ provides the first
compact prediction for the onset of LESA-driven negative ELN flux,
computable from hemisphere-integrated neutrino luminosities and applicable
to any simulation archive producing SH multipole output. Its connection to
fast flavor instability remains a physically motivated hypothesis rather than
a demonstrated consequence of the present data.

\begin{acknowledgments}
The authors thank Thomas Janka (Max-Planck-Institut f\"{u}r Astrophysik,
Garching) for generously providing access to the Garching/Prometheus-Vertex
simulation data and for helpful discussions, and acknowledge the Garching
Core-Collapse Supernova Data Archive as the source of the
Garching/Prometheus-Vertex data used here.
NV acknowledges support from the Millennium Institute of Subatomic Physics
at High Energy Frontier ICN2019\_044.
LJ is supported by a Feynman Fellowship through LANL LDRD project number
20230788PRD1.
\end{acknowledgments}

\section*{Data availability}
\label{sec:data_avail}

The Princeton/Fornax simulation data are publicly available at
\url{https://www.astro.princeton.edu/~burrows/nu-emissions.3d/index.html}.
The Garching/Prometheus-Vertex simulation data are available from the
Garching Core-Collapse Supernova Data Archive~\cite{GarchingArchive} at
\url{https://wwwmpa.mpa-garching.mpg.de/ccsnarchive/}; the corresponding
model publications are cited in Sec.~\ref{sec:garching_data}.

\bibliography{refs_prd_v2}

\end{document}